\documentclass[preprint, aps, pre, eqsecnum,amsmath, amssymb]{revtex4}              
\usepackage[final]{graphicx} 

\begin{document}

\title{Magnetization of rotating ferrofluids: predictions of different
theoretical models}

\author{A.~Leschhorn and M.~L\"{u}cke}
\affiliation{Institut f\"{u}r Theoretische Physik, Universit\"{a}t
des Saarlandes, D-66041~Saarbr\"{u}cken, Germany\\}

\date{\today}

\begin{abstract}

We consider a ferrofluid cylinder, that is rotating with constant rotation
frequency ${\bf \Omega}=\Omega {\bf e}_{z}$ as a rigid body. A homogeneous
magnetic field
${\bf H}_{0}=H_{0} {\bf e}_{x}$ is applied perpendicular to the cylinder axis 
${\bf e}_{z}$. This causes a nonequilibrium situation. Therein the magnetization
 ${\bf M}$ and the internal magnetic field ${\bf H}$ are constant in time and
homogeneous within the ferrofluid. According to the Maxwell equations they are
related to each other via ${\bf H} = {\bf H}_{0} -{\bf M}/2 $. However,
${\bf H}$ and ${\bf M}$ are not parallel to each other and
their directions differ from that of the applied field ${\bf H}_{0}$.
We have analyzed several different theoretical models that provide equations for
the magnetization in such a situation. The magnetization ${\bf M}$ is determined 
for each model as a function of $\Omega$ and $H_{0}$ in a wide range of 
frequencies and fields. Comparisons are made of the different model results 
and the differences in particular of the predictions for the perpendicular
components $H_y =-M_y/2$ of the fields are analyzed.

\end{abstract}

\maketitle                                                                      

\vskip2pc



\section{Introduction}

\begin{figure}[htp] 
\centerline{\includegraphics[width=7cm,angle=0]{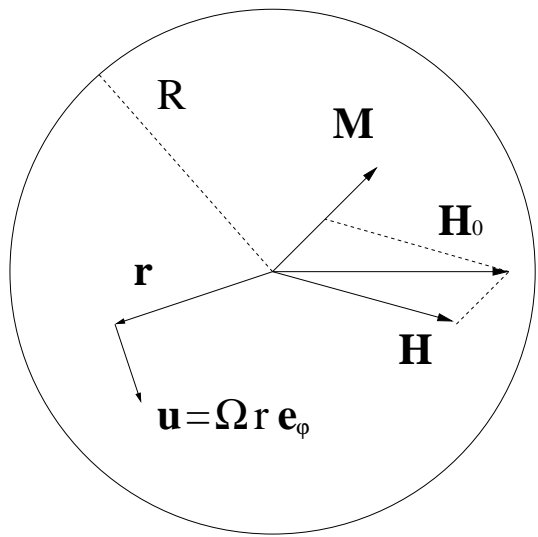}}
\caption{Schematic plot of relevant vectors.}
\label{FIG:vectors}
\end{figure}

There are several theoretical equations for the dynamics of the magnetization 
${\bf M}({\bf r},t)$ of a ferrofluid that is flowing with velocity 
${\bf u}({\bf r},t)$ in an externally applied magnetic field ${\bf H}_0$ 
\cite{ROSEN,S'72,FK,S'01,ML}. Here we compare their predictions for a simple 
special case that is 
experimentally accessible. We consider a ferrofluid cylinder of radius $R$ of 
sufficiently large length to be approximated as infinite in a homogeneous 
applied field ${\bf H}_0 = H_0 {\bf e}_x$ in $x$-direction. The ferrofluid 
cylinder is enforced via its walls to rotate as a rigid-body around its long 
axis with constant rotation 
frequency ${\bf \Omega} = \Omega {\bf e}_z$  being oriented perpendicular to 
${\bf H}_0$. The flowfield is thus 
${\bf u}({\bf r})={\bf \Omega} \times {\bf r} = \Omega r {\bf e}_{\varphi}$ 
where ${\bf e}_{\varphi}$ is the unit vector in azimuthal direction. In such 
a situation all aforementioned models allow for a spatially and 
temporally constant nonequilibrium magnetization ${\bf M}$ that is rotated out
of the directions of ${\bf H}_0$ and ${\bf H}$ by the flow.
The Maxwell equations demand that the fields ${\bf H}$ and ${\bf M}$ within the 
ferrofluid are related to each other via
\begin{eqnarray} 
{\bf H} = {\bf H}_0 - \frac{1}{2} {\bf M}  
\label{pG1}   
\end{eqnarray} 
as indicated schematically in Fig.~\ref{FIG:vectors} and that the magnetic field 
outside the ferrofluid cylinder
\begin{eqnarray} 
{\bf H}^{out} = {\bf H}_0 + \frac{1}{2} \frac{R^{2}}{r^{2}} 
\left( 
2\frac{{\bf r}}{r} \frac{{\bf M}\cdot {\bf r}}{r} 
- {\bf M}   
\right) 
\end{eqnarray}  
 is a superposition of the applied field ${\bf H}_0$ and the dipolar 
 contribution from ${\bf M}$.
 
 \section{Magnetization equations} 

The model equations that we compare here imply a relaxational dynamics either of 
${\bf M}$ towards the equilibrium magnetization 
\begin{eqnarray} 
{\bf M}_{eq} ({\bf H}) =\frac{ M_{eq}(H)}{H} {\bf H} = \chi (H) {\bf H} 
\label{pG2} 
\end{eqnarray} 
or of the "local equilibrium" or "effective" field
\begin{eqnarray}
{\bf H}_{eff} ({\bf M}) =  \frac{M_{eq}^{-1}(M)}{M} {\bf M} = F(M) {\bf M} 
\label{pG3} 
\end{eqnarray}
towards the internal field ${\bf H}$. The equilibrium magnetization 
$ M_{eq}(H)$ referring to the functional relation between internal field $H$ and
magnetization in the case of $\Omega = 0$ is a thermodynamic material property 
of the ferrofluid. The effective field ${\bf H}_{eff}$
lies parallel to ${\bf M}$ and can be seen as the inverse of the defining 
requirement
\begin{eqnarray} 
{\bf M} = {\bf M}_{eq} ({\bf H}_{eff}) \, . 
 \end{eqnarray}  
In equilibrium, $\Omega = 0$, one has ${\bf H}_{eff} = {\bf H}$ and 
${\bf M} = {\bf M}_{eq}$.

We consider here the relations
\begin{eqnarray}  
\mbox{Debye:} \qquad {\bf \Omega} \times {\bf M} 
&=&  
\frac{1}{\tau} \left( {\bf M} - {\bf M}_{eq} \right)   \\
\mbox{S'72 \cite{S'72}:} \qquad {\bf \Omega} \times {\bf M}  
&=& 
\frac{1}{\tau} 
\left( {\bf M} - {\bf M}_{eq} \right) 
+ \frac{\mu_{0}}{4\zeta}  
{\bf M} \times \left( {\bf M} \times {\bf H} \right) \\
\mbox{FK \cite{FK}:} \qquad {\bf \Omega} \times {\bf M}   
&=& 
\gamma_{H} \left({\bf H}_{eff} - {\bf H} \right)     
+ \frac{\mu_{0}}{4\zeta} 
{\bf M} \times \left( {\bf M} \times {\bf H} \right)  \\
\mbox{S'01 \cite{S'01}:} \qquad {\bf \Omega} \times {\bf H}_{eff}    
&=& 
\frac{1}{\tau} 
\left( {\bf H}_{eff} - {\bf H} \right) 
+ \frac{\mu_{0}}{4\zeta} 
{\bf H}_{eff} \times \left( {\bf M} \times {\bf H} \right)  \\
\mbox{ML \cite{ML}:} \qquad {\bf \Omega} \times {\bf M}   
&=& 
\xi ({\bf H}_{eff}-{\bf H})    
\end{eqnarray} 
resulting for the rotating cylinder from the above 5 models. In ML we use the 
weak field variant of ref. \cite{ML}. These equations have to be solved 
numerically in combination with the Maxwell equation (\ref{pG1}).

As an aside we mention that the above equations can be written in the common
form 
\begin{eqnarray}  
{\bf M} \times 
\left( 
{\bf \Omega}
+ \alpha_3  {\bf M} \times {\bf H}_{0} 
\right)    
&=& 
\alpha_1 ({\bf H}_{0} - \alpha_2{\bf M}) 
\label{pG4}  
\end{eqnarray}
with coefficients: 
\begin{tabbing} 
Debye  
\= 
\qquad 
\=
$\alpha_1 = \frac{\chi }{\tau} $ 
\= \quad , \quad \= 
$\alpha_2 = \frac{1}{\chi} + \frac{1}{2}$ 
\= \quad , \quad \=  
$\alpha_3 = 0$ 
\\ 
S'72
\>  \> 
$\alpha_1 = \frac{\chi }{\tau} $ 
\= \quad , \quad \= 
$\alpha_2 = \frac{1}{\chi} + \frac{1}{2}$ 
\> \quad , \> 
$\alpha_3 = \frac{\mu_{0}}{4\zeta}$ 
\\ 
S'01 
\> \> 
$\alpha_1 = \frac{1}{F \tau} $ 
\> \quad , \> 
$\alpha_2 = F + \frac{1}{2} $ 
\> \quad , \> 
$\alpha_3 = \frac{\mu_{0}}{4\zeta}$ 
\\  
FK 
\> \> 
$\alpha_1 = \gamma_{H} $ 
\> \quad , \> 
$\alpha_2 = F + \frac{1}{2} $ 
\> \quad , \> 
$\alpha_3 = \frac{\mu_{0}}{4\zeta}$ 
\\ 
ML 
\> \> 
$\alpha_1 = \xi  $ 
\> \quad , \> 
$\alpha_2 = F + \frac{1}{2} $ 
\> \quad , \> 
$\alpha_3 = 0$ 
\end{tabbing}

\section{Results}
 
\begin{figure}[hbtp] 
\centerline{\includegraphics[width=11cm,angle=0]{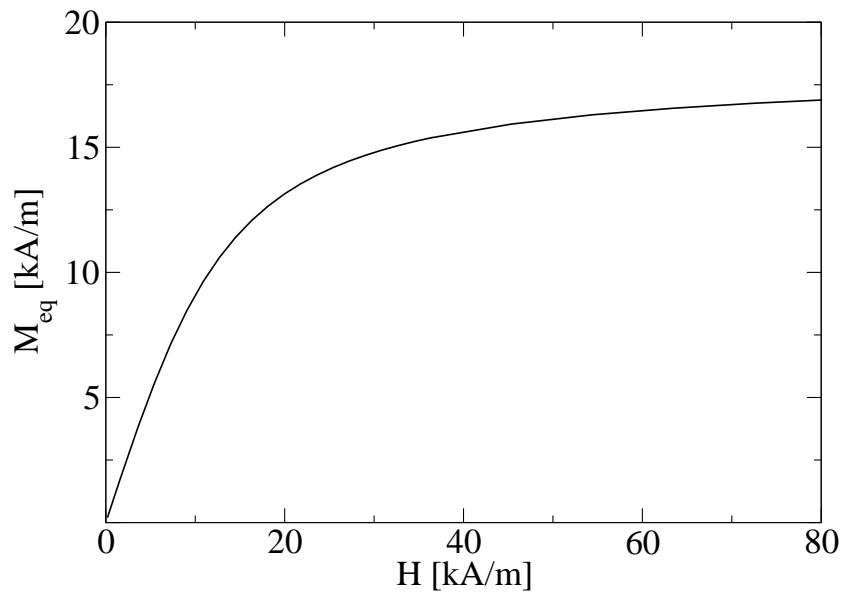}}   
\caption{Equilibrium magnetization $ M_{eq}(H)$ used as input into the models
compared here.}
\label{Meq-H}
\end{figure}

In order to make the comparison of the theoretical results easier we replace
the equilibrium magnetization $M_{eq}(H)$ by the Langevin expression   
$ 
M_{eq}(H) = M_{sat} {\cal L} \left( 3\chi_{0}H/M_{sat} \right) 
$ 
with the initial susceptibility $\chi_{0}=\chi (H=0)$. We use 
$\chi_{0}=1.09$ and $M_{sat}=18149 A/m$ for the saturation
magnetization which is appropriate for the ferrofluid APG 933 of FERROTEC.
The resulting curve is shown in Fig.~\ref{Meq-H}.
Furthermore, we replace the relaxation time $\tau (H)$ by 
$\tau_{B}=6\cdot 10^{-4} s$. For $\zeta \simeq \frac{3}{2} \Phi \eta$ we use the
values $\eta =0.5 Pa\cdot s$ and $\Phi =0.041$ and for $\gamma_{H}$ we
use $\gamma_H =\chi_0 / \tau_B$ \cite{FELD}. For the parameter $\xi$ of ML
\cite{ML} we investigate two different choices: Either the low-field variant, 
$\xi = \chi_0 / \tau_B$, as in FK that is denoted here by ML(F). Or the variant
$\xi =1/[F(M) \tau_B]$ as in S'01 that is denoted here by ML(S).

\begin{figure}
\centerline{\includegraphics[height=12.5cm,angle=0]{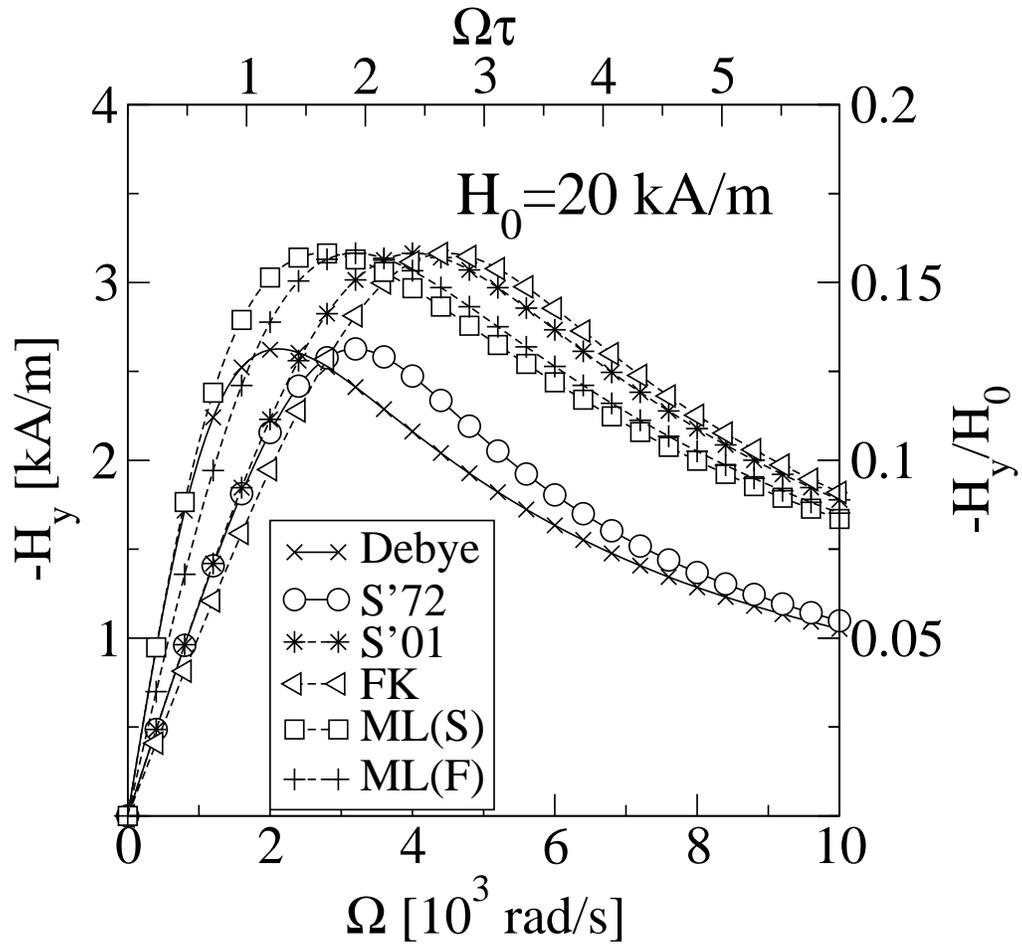}}
\caption{Comparison of the predictions of the different theoretical models for
the transverse internal field $H_y$ versus rotation frequency $\Omega$.}
\label{Hy-Om}
\end{figure}

Especially the perpendicular component $H_{y}=-\frac{1}{2} M_{y}$ of the
magnetic field is suited for a comparison of the different models with each 
other and with experiments. Before doing the former we should like to draw the
attention to the frequency behavior of $M_y(H_0, \Omega)$. We mentioned already
that $M_y$ vanishes for zero vorticity, $\Omega=0$. Furthermore, one finds that 
$M_y$ as well as $M_x$ vanishes also in the limit $\Omega \rightarrow \infty$.
And since one can rewrite the solution of eq.(\ref{pG4}) in the form
$M_{y}= \frac{\Omega \tau}{\alpha_1 + \alpha_3 M^2} \frac{M^{2}}{H_{0}}$ 
one sees that $M_{y}(\Omega)$ has a maximum as a function of $\Omega$ as in Fig.
\ref{Hy-Om}. There we show $H_y$ versus $\Omega$.

\begin{figure}[htp] 
\centerline{\includegraphics[width=12.5cm,angle=0]{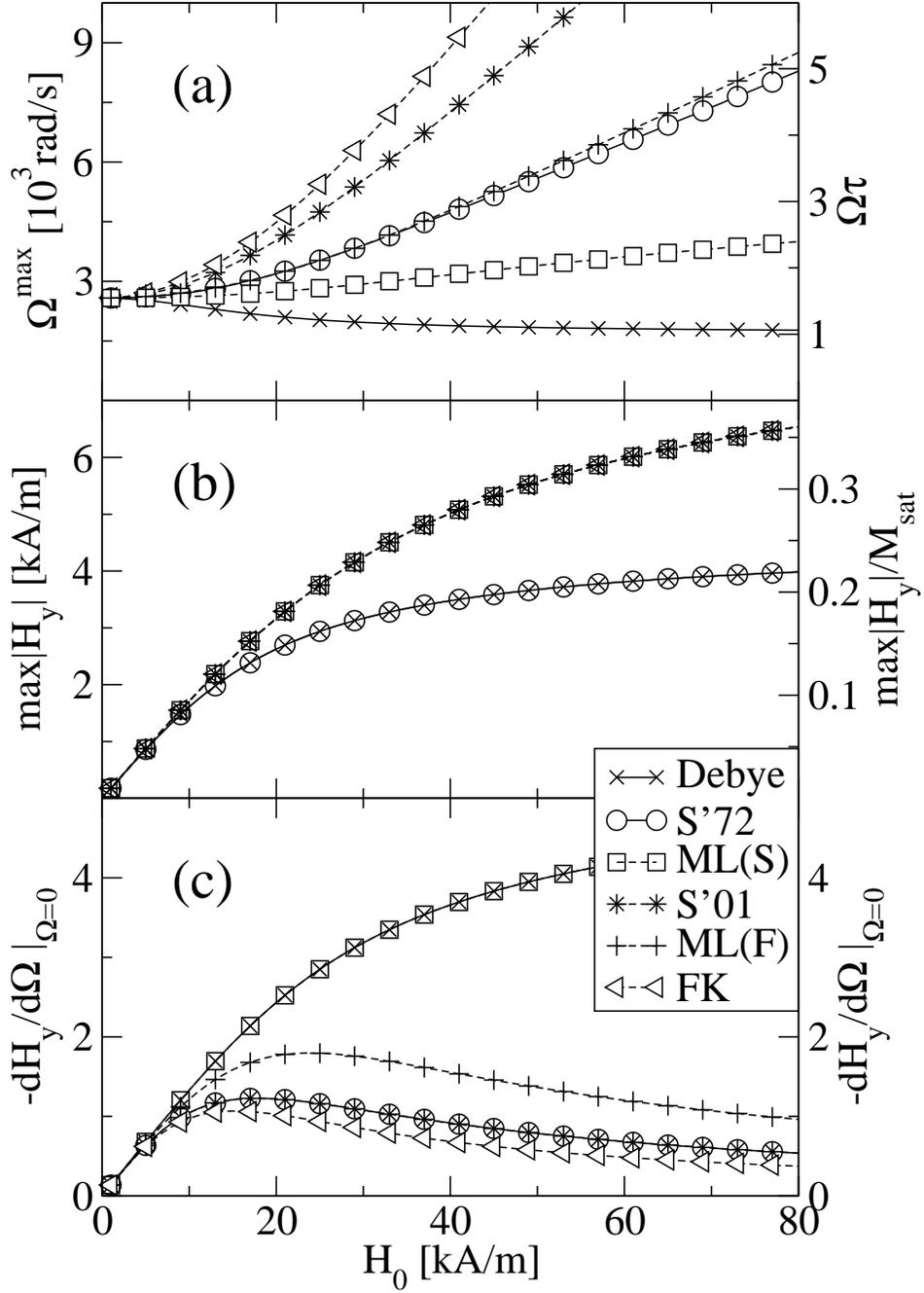}}
 \caption{(a)Frequency $\Omega^{max}$ leading to maximal transverse field, (b)
 largest transverse field, and (c) initial slope $\frac{- d H_y}{d \Omega}$ at 
$\Omega \rightarrow 0$.}
\label{FIG:extrema}  
\end{figure}

The differences in the results for the different models are easily captured
by comparing their
predictions for the maximum values of $|H_y|$, the locations of these maxima at 
$\Omega^{max}$, and the initial slopes $\frac{d |H_y|}{d \Omega}$ at 
$\Omega \rightarrow 0$, each as a function of applied field $H_0$. This is done 
in Fig.~\ref{FIG:extrema}.  

The maximal values of $|H_y|$ of Debye and S'72 are the same while their
locations, $\Omega^{max}$, differ. The models S'01, FK, and ML formulated in
terms of the effective field also share a common maximal value of $|H_y|$ being
larger than that of Debye and S'72 while the location, $\Omega^{max}$, differ
partly substantially. Hence the magnetic torque, ${\bf M}\times {\bf H}$,
entering into S'72, FK, and S'01 only shifts the frequency $\Omega^{max}$.
It remains to be seen whether experiments can be performed with sufficient
accuracy to discriminate between the different theoretical predictions.

\vspace{2cm}

\clearpage
\clearpage
\end{document}